\documentclass[%
 reprint,
 superscriptaddress,
 amsmath,amssymb,
 amsmath,amssymb,
floatfix,
]{revtex4-2}

\usepackage{graphicx}
\usepackage{dcolumn}
\usepackage{bm}
\usepackage{amsmath}
\usepackage[nopatch]{microtype}
\usepackage{booktabs}
\usepackage{physics}
\usepackage{caption}
\usepackage{float}
\usepackage{graphicx}
\usepackage{caption}
\usepackage{subcaption}
\usepackage{amsmath}
\usepackage{caption}
\usepackage{xcolor}
\usepackage{soul}

\begin{document}

\preprint{APS/123-QED}

\title{Reconstruction of quantum states by applying an analytical optimization model}

\author{Rohit Prasad}
\affiliation{Julius-Maximilians-University W\"urzburg,, Technische Physik}
\author{Pratyay Ghosh}
\affiliation{Institute of Physics, Ecole Polytechnique F\'ed\'erale de Lausanne (EPFL), CH-1015 Lausanne, Switzerland}
\affiliation{Julius-Maximilians-University W\"urzburg,, Theoretische Physik I}
\author{Ronny Thomale}
\affiliation{Julius-Maximilians-University W\"urzburg,, Theoretische Physik I}
\author{Tobias Huber-Loyola}
\affiliation{Julius-Maximilians-University W\"urzburg,, Technische Physik}

\date{\today}

\begin{abstract}

When working with quantum states, analysis of the final quantum state generated through probabilistic measurements is essential. This analysis is typically conducted by constructing the density matrix from either partial or full tomography measurements of the quantum state. While full tomography measurement offers the most accurate reconstruction of the density matrix, limited measurements pose challenges for reconstruction algorithms, often resulting in non-physical density matrices with negative eigenvalues. This is often remedied using maximum likelihood estimators, which have a high computing time or by other estimation methods that decrease the reconstructed fidelity. 
In this study, we show that when restricting the measurement sample size, improvement over existing algorithms can be achieved. Our findings underline the multiplicity of solutions in the reconstruction problem, depending upon the generated state and measurement model utilized, thus motivating further research towards identifying optimal algorithms tailored to specific experimental contexts.

\end{abstract}

\maketitle

\section{\label{sec:level1}Introduction}

In quantum physics, we often prepare specific quantum states and verify their preparation by measurement. Measurements always come with noise, where some of the noise might come from the measurement apparatus and some of the noise might come from the physical implementation of the quantum system. To estimate the  quantum state we often prepare the state repeatedly, trying to make perfect copies, which we measure to approximate the system. More specifically, an $n$-qubit quantum system can be fully described by a reconstructed density matrix of dimension  $2^n\times2^n$ by quantum state tomography \cite{Raymer,Leonhardt,Leibfried}, requiring an experiment to measure at least $4^n$ parameters. When restricting measurement time, this exponential scaling leads to a finite number of measurements per parameter. To further minimize the amount of measurement time there have been various approaches such as adaptive measurement techniques and introduction of machine learning for the reconstruction of density matrices\cite{Quek.2021,Deep_learning_2023,Machine_learning_pipeline_2021}. With noisy measurements, a linear reconstruction of the quantum state has a high probability to reconstruct a non-physical density matrix, as it has negative eigenvalues\cite{James2001}.
To remedy this problem, there are several methods \cite{RobinBlumeKohout.2010,PhysRevA.56.1788,PhysRevLett.105.150401}. One approach is to use maximum likelihood estimation, first indroduced by Z. Hradil\cite{Hradil1997}, which has nowadays dozens of implementations, some with direct extremalization\cite{James2001} or iterative solutions\cite{Rehacek2001,FMLE}. However, we want to focus on three widely used implementations for a full quantum state tomography (QST) and to estimate the physical density matrix using maximum likelihood. Namely maximum likelihood estimation as introduced by James in 2001 (MLE) \cite{James2001}, iterative maximum likelihood estimation as introduced by Je{\v{z}}ek et al. in 2003 (iMLE) \cite{FMLE}, and an algorithm proposed by J. A. Smolin, J. M. Gambetta, and G. Smith (SGS)\cite{ibm}. While MLE and iMLE are the best known algorithms for getting the highest possible fidelity, when comparing to a known quantum state, SGS has a very fast runtime, with a slight reduction of reconstructed state fidelity. SGS has been demonstrated to reconstruct quantum states with up to 14 qubits within reasonable runtime\cite{Hou2016}. We want to point out, that MLE can be sped up as well using the accelerated projected-gradient method, but it is still not as fast as SGS\cite{Shang2017}. 
We took the approach of SGS as a role model, the fastest correction algorithm and we refined their algorithm using optimization on a simpler log-likelihood. This gives improvements in the regime where we restrict the sample size as detailed below.

 In this paper, we investigate the reconstruction of the density matrix, using MLE, iMLE, SGS and our new algorithm, and compare their performance for a finite measurement time with reasonable measurement samples of $10^3$ counts for each element in the basis set($4^n$). We are using a sample size of $10^3$, since it was shown\cite{FMLE} that MLE yields inaccurate results for very small number of detection events and we want to use MLE as a benchmark. Our approach results in an algorithm with better fidelity compared to SGS while keeping the algorithm runtime exactly same , however, the fidelity is slightly underestimated as compared to MLE and iMLE. 

\section{Reconstruction and Cost function}

A $n$-qubit state is mathematically represented by a $2^n\times2^n$ dimensional density matrix.
For a full reconstruction of the density matrix a basis set of $4^n$ orthonormal hermitian matrices $\hat{\Gamma}$ is needed \cite{James2001}.
One such basis set is the tensor product of all permutations of the Pauli matrices. For example, for a 2-qubit system the basis set is formulated as $\hat{\sigma}_i \otimes \hat{\sigma}_j$ ($i,j = 0,1,2,3$) \cite{chuang}, where $\hat{\sigma}_{i,j}$ are the Pauli matrices. 
Even after employing full tomography, the linear-reconstructed  density matrix $\rho_r$ is often non-physical. The non-physicality persists even if you follow more advanced protocols such as partial tomography and adaptive tomography\cite{Adaptive_quantum_state_tomography_2017}, since it comes from the linear reconstruction.
To reconstruct the most accurate approximation of the physical density matrix $\rho_p$ characterized by all positive eigenvalues, MLE uses maximization of likelihood between the potential density matrix and acquired measurement data. Since the known algorithms for minimization are better researched than for maximization, the problem is usually rewritten by taking the logarithm of the likelihood function and applying a negative unit multiplication. This transforms the objective into the minimization of log-likelihood\cite{James2001,ibm},
\begin{equation*}
     \mathcal{L}_{log,MLE} = \sum_{j}^{4^n}  \frac{[ \bra{m_j} \rho_p \ket{m_j} - m_j ]^2}{2 \bra{m_j} \rho_p \ket{m_j}}
\end{equation*}
\\
Here, $\ket{m_j}$ and $m_j$ are the measurement basis state and respective state's probability. While this reconstruction gives very good results, it is also computationally expensive, because the minimization has $4^n$ summands.  
To make it faster, while still keeping positive eigenvalues the SGS algorithm. After linear reconstruction, the algorithm follows 3 simple steps: Step 1: Nullify all the negative eigenvalues, while keeping the sum of the negative eigenvalues as `$a$'. Step~2:~Dividing and adding `$a$' equally to all the positive eigenvalues. This step ensures that $\mathrm{Tr}(\rho_p)=1$, which can be understood as normalization of the density matrix. Step 3: If this results in new negative eigenvalues, repeat steps 1 and 2. With all negative eigenvalues eliminated, we can proceed construction of the physical density matrix  $\rho_p =  \sum_{i}^{2^n} \lambda_i \ket{v_i}\bra{v_i}$, where $\lambda_i$ is the ith eigenvalue as found above and $\ket{v_i}$ is its corresponding eigenvector.
While this method effectively addresses non-physical eigenvalues, it does not result in the best possible fidelity for a given measurement sample size, as can be seen by comparing it to MLE, see Fig.~\ref{fig:ivc}.

\section{Eigenvalue Optimization Algorithm}

We found that the SGS algorithm can be improved by integrating problem specific optimization techniques. Our method follows the steps of SGS, but instead of evenly distributing the negative eigenvalues, we modify the distribution with a weighted normalization.
To get the weighted function, we reformulate the cost function into a simplified expression:
\begin{equation}\label{eq:full_cost}
\begin{split}
     \mathcal{C} =\sum_{j}^{4^n} \Bigr[ &\bra{m_j} \rho_p \ket{m_j} - m_j \Bigr]^2\\
     =\sum_{j}^{4^n} \Bigr[ &\bra{m_j} \rho_p \ket{m_j} - \bra{m_j} \rho_r \ket{m_j} \Bigr]^2\\
     =\sum_{j}^{4^n}  \biggr[& \bra{ m_j} \Bigr[\sum_{i}^{2^n} \lambda_i \ket{v_i}\bra{v_i} \Bigr]\ket{m_j} \\
     &-\bra{m_j} \Bigr[\sum_{i}^{2^n} v_i \ket{v_i}\bra{v_i}\Bigr] \ket{m_j} \biggr]^2\\
    =\sum_{j}^{4^n}\Bigr[&\sum_{i}^{2^n} (\lambda_i-v_i)\bigr|\bra{v_i} \ket{m_j}\bigr|^2\Bigr]^2\\
    =\sum_{j}^{4^n}\Bigr[&\sum_{i}^{2^n} d_i\bigr|\bra{v_i} \ket{m_j}\bigr|^2\Bigr]^2
\end{split}
\end{equation}
\\
with $v_i$,$\ket{v_i}$ as the eigenvalue and eigenvector of a linearly-reconstructed density matrix $\rho_r$ and $\lambda_i$,$\ket{v_i}$ as the eigenvalues and eigenvectors of the target physical density matrix $\rho_p$. The eigenvalues $v_i$ are arranged in descending order.
Under the assumption that all negative eigenvalues are set to zero, the cost function is further simplified. Rewriting the cost function introducing $d_i = \lambda_i - v_i$ and $n_+$ as the count of non-negative eigenvalues: 
\begin{equation}\label{eq:cost_eo}
\begin{split}
     \mathcal{C} &= \sum_{j}^{4^n}\Bigr[\sum_{i = 1}^{n_+} d_i\bigr|\bra{v_i} \ket{m_j}\bigr|^2+c_0\Bigr]^2
     \\
     c_0 &=  -\sum_{i = n_++1}^{2^n} v_i\bigr|\bra{v_i} \ket{m_j}\bigr|^2
\end{split}
\end{equation}
\\
\begin{figure}[h]
	\centering
		\includegraphics[width=0.47\textwidth]{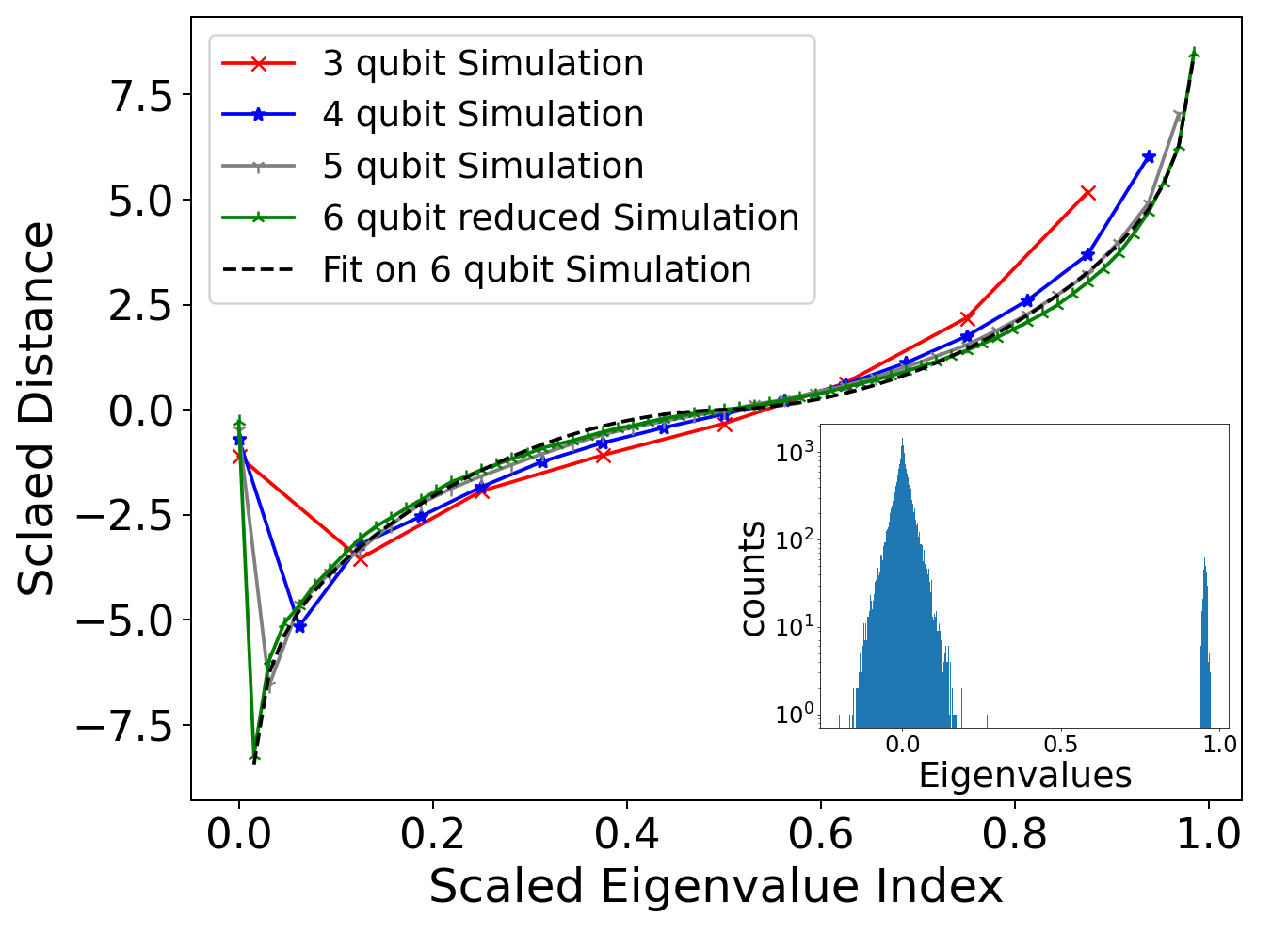}
		\caption{\textbf{Scaled distance vs Scaled eigenvalue index.} Distance ($\lambda_i-v_i$) has been represented with a upscaling of $2^n$ and eigenvalue index with a scaling down of $1/2^n$ to make all the qubits curve overlap. Each curve has been averaged over 1000 optimized $d_i$ of random density metrics. The fit curve represents an odd power series optimized to fit the scaled distance vs. scaled eigenvalue index curve of 6 qubit curve. The inset is a histogram graph of the eigenvalues calculated from the non physical reconstructed density matrix of the random state used in the simulation.}
  \label{fig:fit}
\end{figure}
For $\Sigma \lambda_i = \Sigma v_i = 1$ we have $\Sigma d_i = 0$. As per the assumptions $d_i = -v_i$ for $i>n_+$, to get $\Sigma d_i = 0$ we get a constraint that is $\Sigma_1^{n_+} d_i = \Sigma_{n_++1}^{2^n} v_i$, which constricts $d_i$'s from evaluating to zero for minima. This constrain has been achieved defining  $d_{n_+}= -\Sigma_1^{n_+-1} d_i+ \Sigma_{n_++1}^{2^n} v_i$. This leaves ($n_+-1$) $d_i$'s for the minimization of cost function (Eq.~\ref{eq:cost_eo}).

Using Constrained Optimization by Linear Approximation (COBYLA) with constraints $d_i\geq-v_i$ to ensure that the final eigenvalues are greater than zero($\lambda_i\geq 0$), the cost function has been minimized\cite{Powell_1998,Powell1994ADS,Powellchap}.
Fig.~\ref{fig:fit} represents the plot between $d_{i,\mathrm{norm}}= d_i\cdot2^n/\sum_{n_++1}^{2^n} |d_i|$ in y-axis against normalized index $i$ between 0-1 in the x axis. The plot is generated by simulating the defined protocol on a 3-, 4-, 5- and 6- qubit system. 
The simulation has been performed for a total measurement number of $10^3\times4^n$ and averaged over $10^3$ random states with a varying purity between $0.5$ to $1$.
We observe that the maximum eigenvalue, typically close to 1, regarded as the eigenvalue of the primary eigenvector, has $d_i \approx 0$ , regardless of the system's size. With this observation, we set $d_1 = 0$ for our further analysis.
The figure also includes a fit curve for $d_i$ (excluding $d_1$) of the 6-qubit system, as there is sufficient data available from the 6-qubit system to extrapolate to an n-qubit system. In other words, the distribution function can be used for arbitrarily big qubit systems without the need to run optimization on these big systems.
\\\vspace{1pt}\\
\textbf{Physical matrix reconstruction}
\\
\vspace{-5pt}
\\
We introduce the Eigenvalue Optimization (EO) algorithm, which utilizes the above found weighted normalization curve, see Fig.~\ref{fig:fit}, to construct the physical eigenvalues from the linearly-reconstructed density matrix. To obtain the weighted normalization from the fit curve, we adjusted the fit curve to scale between 0 and 2. This adjustment enables us to set the function value to 0 at 1. Moving forward, we define the fit function as follows,
 
\begin{equation}\label{eq:4}
     F(x) = \sum_{k=1}^{6} c_k (x-1)^{(2k-1)},
\end{equation}
where $c_1 = 1.21$, $c_2~=~21.03$,
  $c_3~=~-119.80$, $c_4~=~339.23$,
 $c_5~=~-418.31$, $c_6~=~188.26$, was found through fitting. The function fits the data with $\chi^2 < O(10^{-3})$. The choice of $\chi^2 < O(10^{-3})$ is random. If more error is acceptable, the function could be fitted with fewer power series terms. \\

 To construct the physical eigenvalues starting from the non-physical density matrix the following steps are required to be followed:
\begin{enumerate}
   \item Calculate the eigenvalue $v_i$ and eigenvector $\ket{v_i}$ from the reconstructed density matrix $\rho_r$, arranged in descending order.
   \item Define the sum of negative eigenvalues $a_{mod}$, a normalization $N_f$ for compact writing:
   \vspace{10pt}\\
   $a_{mod} = \sum_{n_++1}^{2^n}|v_i|$ 
   \vspace{10pt}\\
   $N_f = \sum_{i=2}^{n_+} F(\frac{i-1}{n_+-0.5})$
   
   \item Define the new eigenvalues $\lambda_i$ according to
   \vspace{10pt}\\
   $
    \lambda_i= \begin{cases}
        v_1 &i = 1\\
             v_i + \frac{a_{mod}}{N_f}\cdot F(\frac{i-1}{n_+-0.5})  &i\in [2,n_+]\\
            0  &i\in [n_++1,2^n]
    \end{cases}
    $
    
    \item If after these steps there are still some negative eigenvalues, i.e.,  $\lambda_i < 0$, steps 2 and 3 are repeated. For step 2 the old non-physical eigenvalues $v_i$ are replaced by $\lambda_i$ found in step 3.
    \item Now all eigenvalues are positive, resulting in a physical matrix \\
    $\rho_p =  \sum_{i}^{2^n} \lambda_i \ket{v_i}\bra{v_i}$
\end{enumerate}

\section{Simulation}

We conducted simulations on 1000 random 2-qubit density matrices. These simulations involved varying the total number of coincidence counts $\mathcal{N}$ over cube basis measurement\cite{choice_of_measurment_2008} with added Gaussian error. We approximate the underlying multinomial distribution by a Gaussian error in each probabilistic outcome, because of the central limit theorem.

\begin{figure}[hb]
	\centering
		\includegraphics[width=0.47\textwidth]{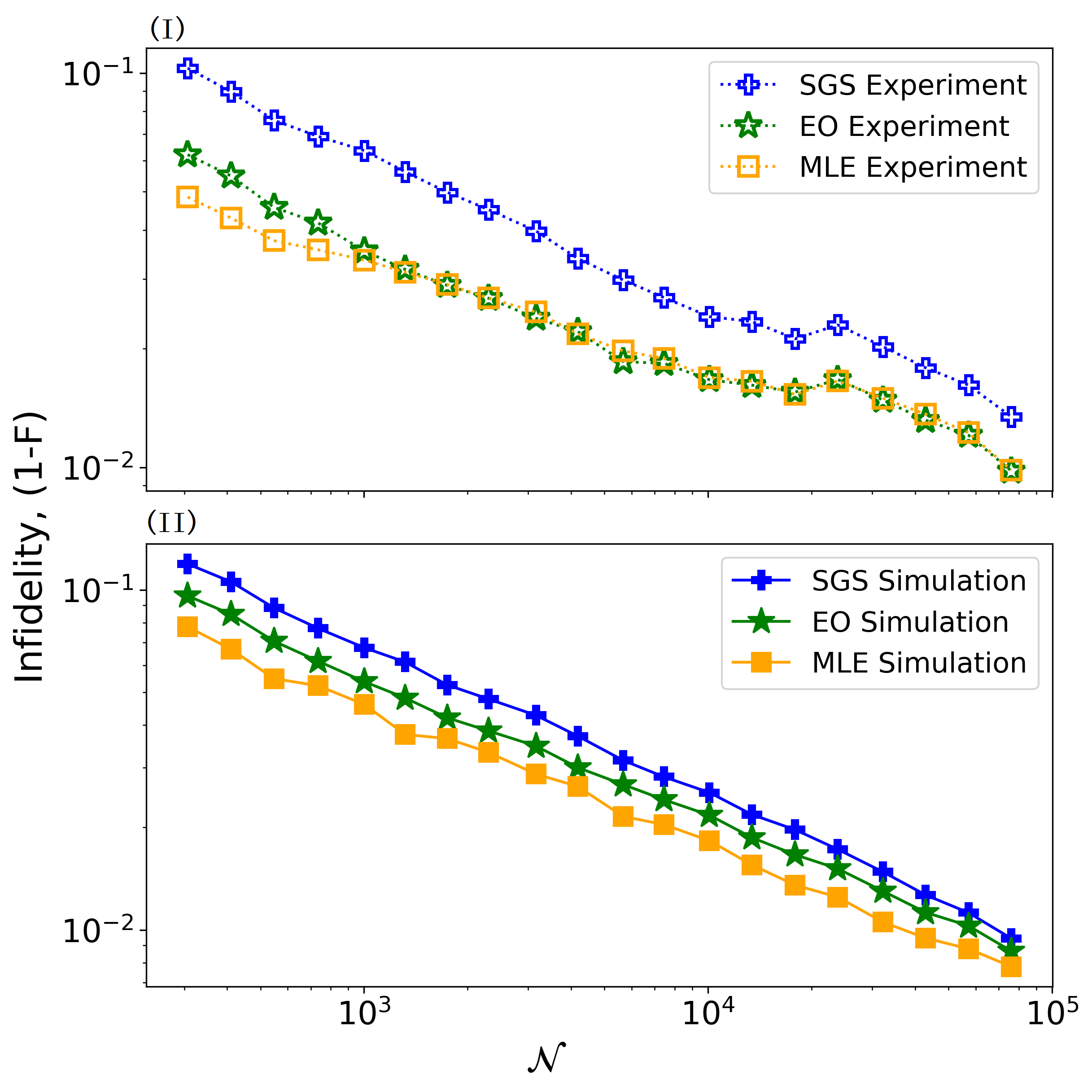}
 \caption{Simulated and Experimental results of \textbf{Infidelity vs. Total coincidence counts($\mathcal{N}$)}. \textbf{(I)}  We performed the 2-qubit reconstruction experiment of IBM's Manila Quantum computer to get a real data set from a NISQ device. Each point has been averaged over 100 experiments. The purity of the quantum state in the experiment has been calculated to be $Tr\{\rho^2\}$ = 0.94. \textbf{(II)} Simulation of a 2 qubit quantum state with the same purity. The simulation has been performed to compare the infidelity of the reconstructed density matrix, with each point being averaged over 1000 random 2-qubit quantum states}
 \label{fig:ivc}
\end{figure}

The mixed state was generated by adding the identity matrix and a rotation error to a pure state. Mathematically, this process is represented as 
\begin{equation}
    \rho_{mixed} = p\cdot\rho_{pure} + \frac{1-p}{3}\cdot(2\cdot \textbf{I}_{norm} +\textbf{O}_{error} \cdot\rho_{pure} ),
\end{equation}where $p$ represents the purity parameter, $\rho_{pure}$ is the initial pure state, $\textbf{I}_{norm}$ denotes the normalized identity matrix, $\textbf{O}_{error}$ and is the rotation error matrix.
This mixed state exhibits all positive and non-zero eigenvalues, leading to a high proportion of reconstructed quantum states being physical even with a limited set of measurements. 
This is usually not the case for experimental data. Thus, to emulate experimental data accurately, we have set a fraction of eigenvalues to be zero and added an gaussian error to the projection measurements. 
The purity value of 0.94 was selected because it matches the purity of the experimental data used for comparison. Specifically, an experiment was conducted on the IBM quantum computer ``Manila"\cite{IBM_Manila}, utilizing the Bell state $\ket{\psi^+} = (\ket{10}+\ket{01})/2$, cube measurement basis, and 100 samples for each value of $\mathcal{N}$.

\begin{figure}[h]
	\centering
\includegraphics[width=0.47\textwidth]{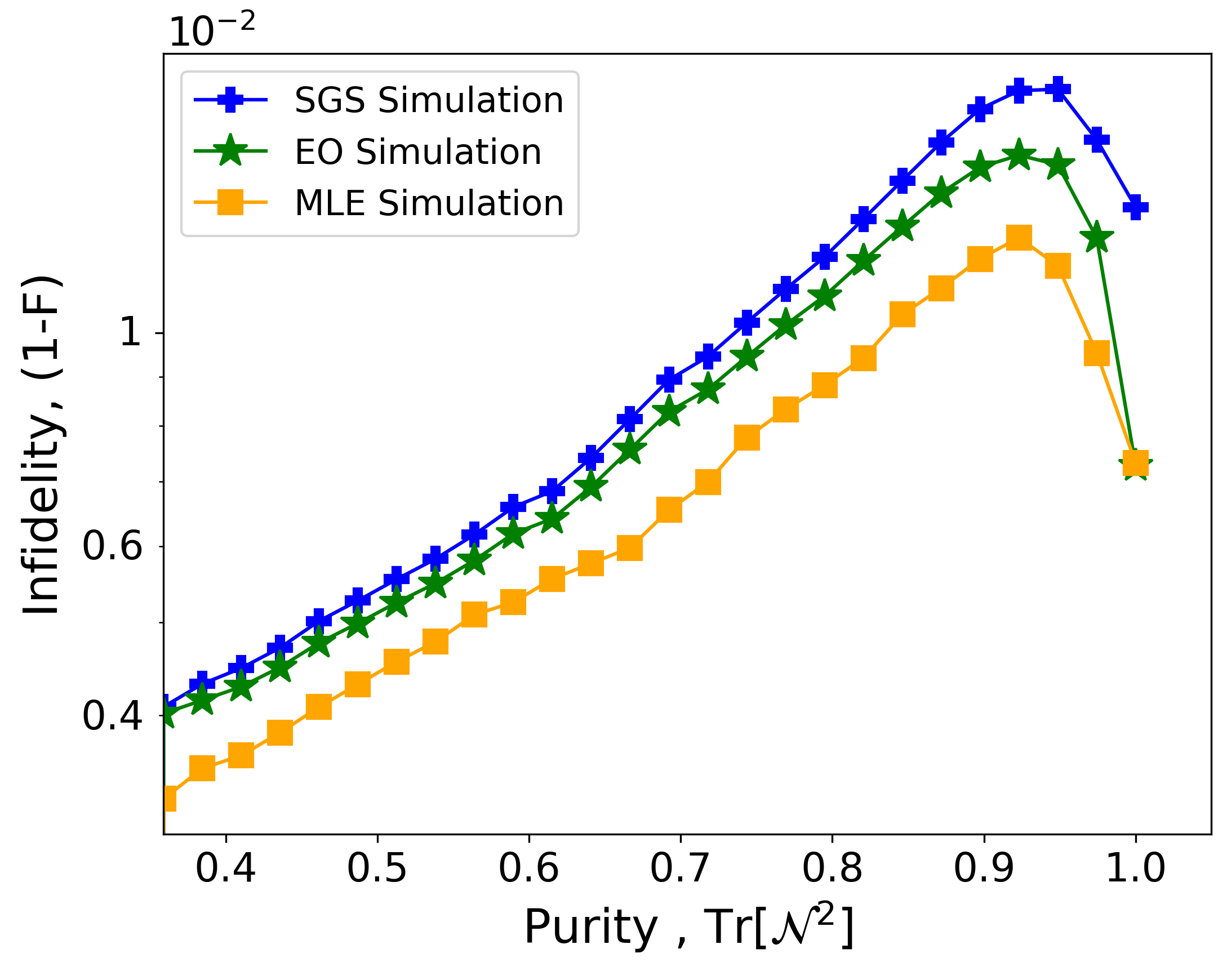}

 \caption{\textbf{Simulation of infidelity vs Purity of density matrix.} Keeping the total number of counts $\mathcal{N} = 4^2 \times10^3$ each point has been averaged over 1000 randomly simulated 2 qubit states}
 \label{fig:ivp}
\end{figure}

In the simulations, the infidelity of $\rho_p$ is calculated with the ideal density matrix, as this matrix is already known.  However, in the case of the experimental data, the ideal density matrix is not available. Therefore, we use a reconstructed density matrix obtained from a comparatively high number of counts, which is $\mathcal{N} = 10^7$, in place of the ideal density matrix. Calculating  $\text{Infidelity}(\rho,\sigma)= 1-\text{Fidelity}(\rho,\sigma)$ $=1-Tr^2\Bigr(\sqrt{\sqrt{\rho}\sigma\sqrt{\rho}} \Bigr)$ to compare algorithms for simulated and experimental data has an intuitive appeal, as it reflects the error reduction as the total number of measurements increases.\\
Fig.\ref{fig:ivc} depicts the dependence of infidelity on the number of counts. The clear similarities between the trends in infidelity for both the experimental and the simulated data indicate consistency and support the validity of our error-induced data model. We compare the performance of SGS, EO, and MLE for experimental data, obtained from IBM's Manila and for simulated data. While the SGS algorithm provides a good approximation of the minimum log-likelihood, it is evident that the improved algorithm calculates a density matrix that is closer in fidelity to the density matrix constructed using the MLE. The MLE is presumed to offer the highest fidelity attainable with a given set of data.

The purity of the quantum system influences the performance of a reconstruction algorithm. Thus, we show the performance of EO Fig.~\ref{fig:ivp} compared to the performance of MLE and SGS for different purities, with EO being closer to MLE for all purity values. We find it particularly interesting that the infidelity is maximal not at a state purity of one. It is quite intuitive that when adding more mixed states into the total state, i.e. assuming a state with lower purity, that the infidelity of the reconstruction is small, since the linear reconstructed state is often already physical. However, the estimation infidelity peaks at around a purity of 0.9, which has also been reported elsewhere and was attributed to a misestimation of small eigenvalues~\cite{Adaptive_quantum_state_tomography_2017}.
\begin{figure}[h]
  \begin{subfigure}[b]{0.4\textwidth}
    \includegraphics[width=\textwidth]{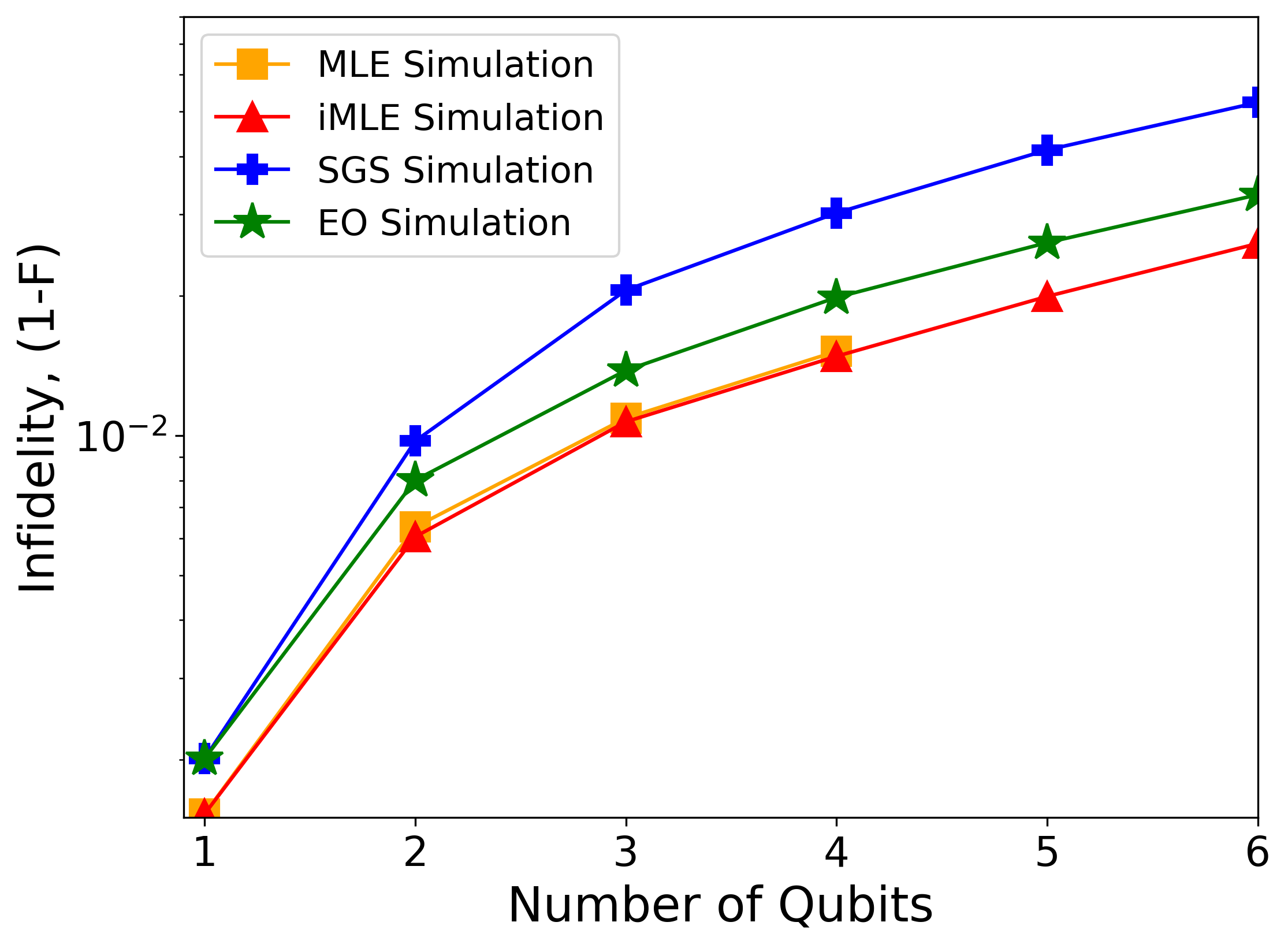}
    \caption{Infidelity vs number of qubits}
    \label{fig:ivn}
  \end{subfigure}
  \hfill
  \begin{subfigure}[b]{0.4\textwidth}
    \includegraphics[width=\textwidth]{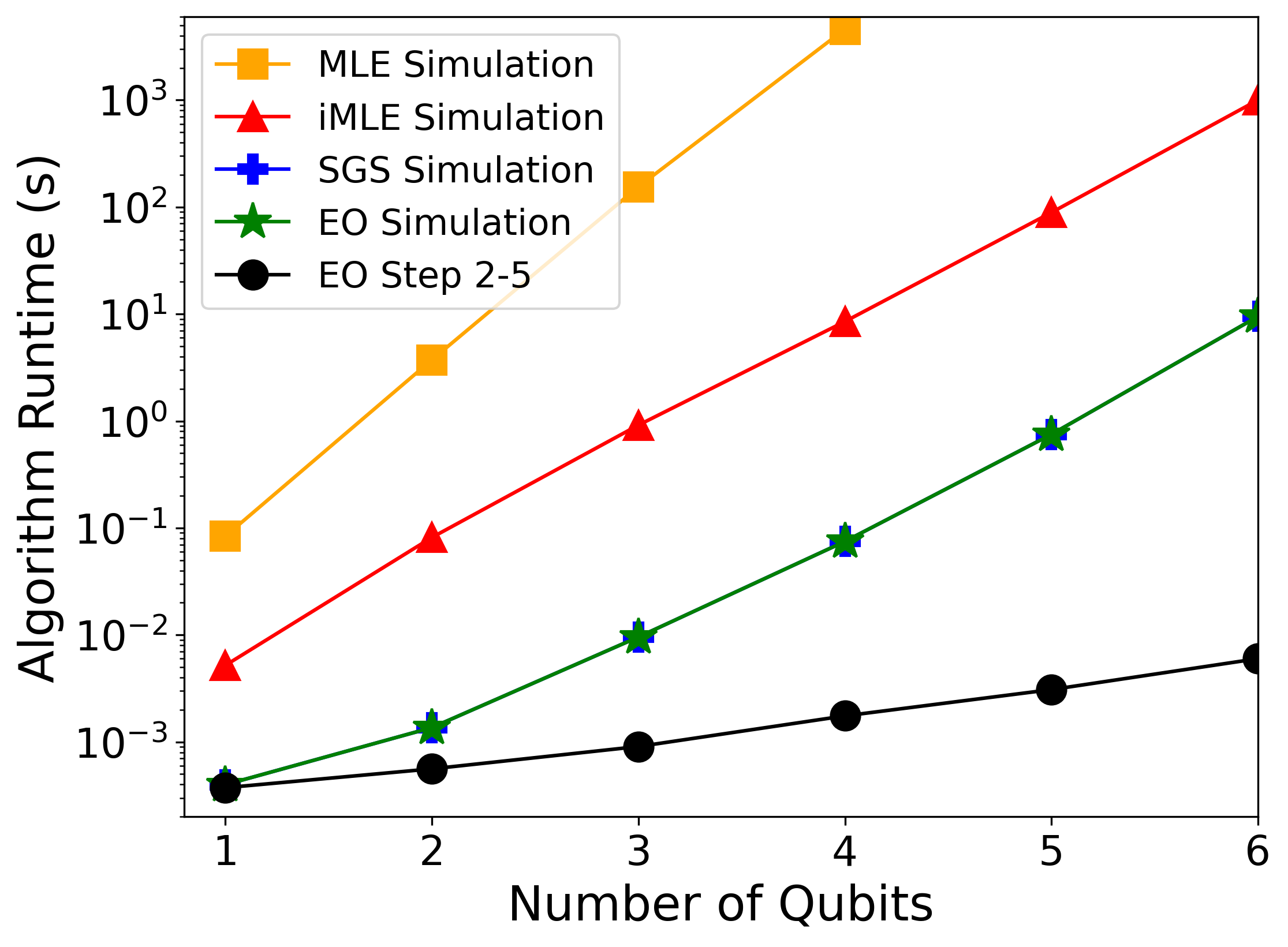}
    \caption{Algorithm runtime vs number of qubits}
    \label{fig:tvn}
  \end{subfigure}
  \caption{ \textbf{Simulation of infidelity vs number of qubit.} \textbf{a.} With a fixed total number of counts $\mathcal{N} = 4^n \times10^3$ the plot shows the infidelity of each method vs. the number of qubit. \textbf{b.} The plot presents the run-time of all algorithms vs. the number of qubit. The timing has been performed in Python on a single core of an Intel i5- 7500 CPU running at 3.4GHz. Each point averaged over 1000 random metrics, except for the 4-qubit MLE case, which was averaged over 100 random matrices}
\end{figure}

While a 2-qubit system is already interesting, the performance of reconstruction algorithms becomes particularly compelling for larger systems, given that the size of the Hilbert space scales exponentially. Thus, we perform calculations for systems up to 6 qubits. 
Fig.~\ref{fig:ivn} compares the infidelity of the density matrix given by the algorithms for different system sizes. For each point in the plot, the infidelity has been averaged over 1000 randomly simulated $n$-qubit states, except for the 4-qubit MLE case, which was averaged over 100 random matrices, with a fixed number of counts $\mathcal{N} = 4^n\times10^3$. We can see an increase in the infidelity of the corrected density matrix as we are working with a fixed limited set of $10^3$ measurements performed for each of $4^n$ projection operators, which increases the overall error, as the number of projections increase. For a single measurement the error changes with increased counts $\mathcal{N}$ roughly as the expected $\frac{1}{\sqrt{\mathcal{N}}}$, compare to Fig.~\ref{fig:ivc}. Additionally, from the same simulation, the runtime of the three algorithms has been plotted, as shown in the Fig.~\ref{fig:tvn}. We were not able to perform simulations for MLE for 5 and 6 qubits on a desktop PC, since the runtime for a single simulation for 4 qubits already took over 1000 seconds, as seen in the Fig. 4b.
Both the SGS and EO algorithms operate on the same principle of computing eigenvalues and eigenvectors of a density matrix to reconstruct the physical eigenvalues. To confirm whether the data has resulted in a physical or a non-physical density matrix one needs to compute the eigenvalues and eigenvectors prior to running the correction algorithm. For the complete reconstruction algorithm this time has to be considered and results in a computational complexity of order $O(d^3)$ (where $d$ is the dimensionality of the Hilbert space). However, if one considers only steps 2-5 of the algorithm, i.e., already knowing the eigenvalues from the prior check done for physicality, the runtime needed by EO is significantly faster, see black circles in the Fig.~\ref{fig:tvn} to highlight the actual time advantage of the algorithm. If one can find a faster algorithm for finding the eigenvalues, the EO algorithm can give the plotted time advantage. The same argument holds for SGS, since the two algorithms exhibit the same computational time and possess an exponential time advantage over MLE \cite{ibm,eigenval_1}. Notably, a clear advantage is observed for EO over SGS, as EO achieves lower infidelity while maintaining identical runtime as SGS for the selected range of measurements.

\section{Conclusion}
We have introduced the Eigenvalue Optimization (EO) algorithm to construct the physical density matrix by nullifying negative eigenvalues and re-normalizing positive eigenvalues. Our algorithm provides an analytical solution derived from an improved version of a simpler log-likelihood cost function (Eq. \ref{eq:cost_eo}) obtained via numerical optimization. We provide a fit to the distribution function, enabling to use our method without redoing the optimization and with the possible extension to larger qubit systems.
Notably, our algorithm achieves the same runtime as the SGS algorithm and is exponentially faster than the MLE\cite{James2001} and iMLE\cite{FMLE}. As EO and SGS uses modification of eigenvalue in the order of runtime $O(2^n)$ (EO Steps 2-5 Fig.~\ref{fig:tvn}). Theoretically, it can be asserted that this is the fastest algorithm, as just achieving the coverage of all eigenvalues necessitates a minimum runtime of $O(2^n)$.
Moreover, it outperforms the SGS algorithm across a wide range of purities, considering a limited set of $10^3\times 4^n$ measurements.
Our findings represent a general reconstruction framework for transforming non-physical density matrices into physical ones, anticipating the emergence of weighted normalization schemes tailored to specific experimental protocols. They underline the multiplicity of solutions in the reconstruction problem, depending upon the generated state and measurement model utilized, thus motivating further research towards identifying optimal algorithms tailored to specific experimental contexts.
\section{Acknowledgements}
The Julius-Maximilians-Universität Würzburg is grateful for financial support and funding for this work from the German Ministry for Education and Research (BMBF) through the project QECS (FKZ: 13N16272) and R.T. acknowledges support from the DFG through QUAST FOR 5249-449872909 (Project P3), through Project-ID 258499086-SFB 1170, and from the Würzburg-Dresden Cluster of Excellence on Complexity and Topology in Quantum Matter-ct.qmat Project-ID 390858490-EXC 2147.
We are also grateful to Dr. Miroslav Je\v{z}ek for insightful advice and helpful discussion.

\nocite{*}

\bibliography{apssamp}

\end{document}